\documentclass[final,3p,times,twocolumn]{elsarticle}

\usepackage{graphics}
\usepackage{graphicx}
\usepackage{amssymb}
\usepackage{amsthm}
\usepackage{amsmath}
\usepackage{upgreek}
\usepackage{subfigure}
\usepackage{wasysym}
\usepackage{textcomp}
\usepackage{booktabs}
\usepackage{paralist}

\biboptions{square}
\journal {Nuclear Instruments and Methods in Physics Research Section A}

\begin{document}
\begin{frontmatter}

\title{Design and implementation of wire tension measurement system for MWPCs used in the STAR iTPC upgrade}

\cortext[cauthor]{Corresponding autor. Tel.: +86 531 88364515.\\
Address: 27 Shanda South Road, Jinan, Shandong, China, 250100\\
E-mail: xuqh@sdu.edu.cn}

\author[1:sdu]{Xu Wang}  
\author[1:sdu]{Fuwang Shen}  
\author[1:sdu]{Shuai Wang}  
\author[1:sdu]{Cunfeng Feng}  
\author[1:sdu]{Changyu Li}   
\author[1:sdu]{Peng Lu}     
\author[2:lbl]{Jim Thomas}  
\author[1:sdu]{Qinghua Xu\corref{cauthor}}   
\author[1:sdu]{Chengguang Zhu}  

\address[1:sdu]{School of Physics and Key Laboratory of Particle Physics and Particle Irradiation (MOE), Shandong University, Jinan 250100, China}
\address[2:lbl]{ Nuclear Science Division, Lawrence Berkeley National Laboratory, Berkeley, CA94720, USA}

\begin{abstract}
The STAR experiment at RHIC is planning to upgrade the Time Projection Chamber which lies at the heart of the detector. We have designed an instrument to measure the tension of the wires in the multi-wire proportional chambers (MWPCs) which will be used in the TPC upgrade. The wire tension measurement system causes the wires to vibrate and then it measures the fundamental frequency of the oscillation via a laser based optical platform. The platform can scan the entire wire plane, automatically, in a single run and obtain the wire tension on each wire with high precision. In this paper, the details about the measurement method and the system setup will be described. In addition, the test results for a prototype MWPC to be used in the STAR-iTPC upgrade will be presented.
\end{abstract}


\end{frontmatter}

\section{Introduction}

The STAR detector is located at the Relativistic Heavy Ion Collider (RHIC) at Brookhaven National Laboratory (BNL). It uses a Time Projection Chamber (TPC) as its primary tracking device\cite{ref:detector_1,ref:detector_2,ref:detector_3,ref:detector_4,ref:detector_5}. The TPC records the tracks of particles, measures their momenta in a 0.5 T magnetic field, and identifies the particles by measuring their ionization energy loss (dE/dx). Its acceptance covers a pseudo-rapidity range of $|\eta|$ $\textless$ 1 over the full azimuth. Particles are identified over a transverse momentum range from 100 MeV/c to greater than 1 GeV/c, and momenta are measured over a range of 100 MeV/c to 30 GeV/c. \\

The STAR TPC has played a central role in the RHIC physics program for over 15 years. STAR has decided to upgrade the inner sectors of the TPC (iTPC) so that the pseudo-rapidity coverage of TPC will be extended from $|\eta|$ $\textless$ 1 to $|\eta|$ $\textless$ 1.5 with improved energy loss dE/dx measurements for particle identification and better tracking performance. The iTPC upgrade will require new readout electronics to match the increased number of read-out channels in the inner sectors. The upgrade will also replace the wire grids in the MWPC (multi-wire proportional chamber) readout system so they can be run at lower gain and utilize larger pads. The new wire grids will extend the lifetime of the STAR TPC into the next decade and the increased acceptance of the new pad-planes will allow STAR to pursue an enhanced physics program in the Beam Energy Scan II program and beyond.\\

\begin{figure}[hbtp]
\begin{center}
\includegraphics[width=0.45\textwidth]{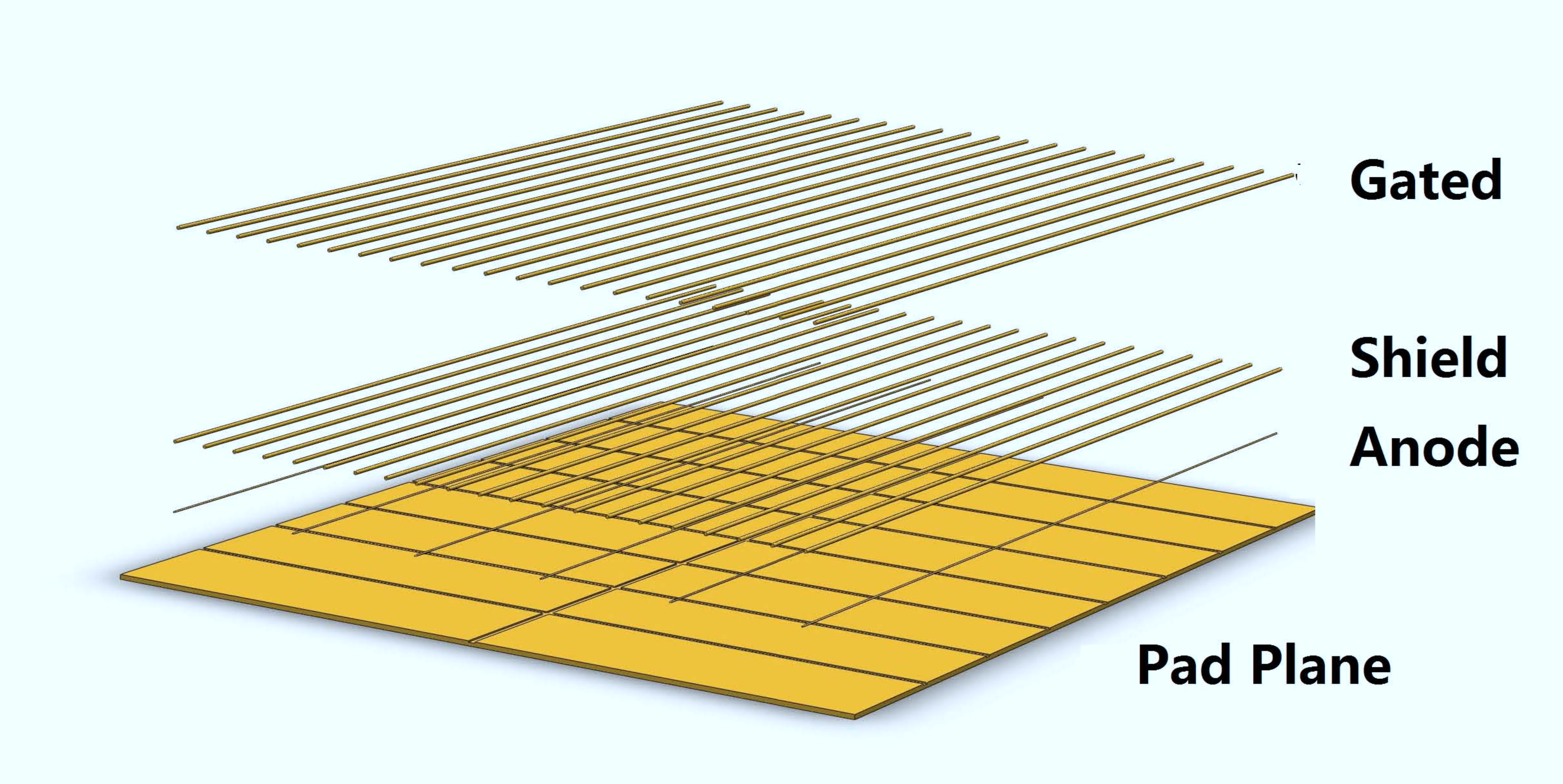}
\caption{A schematic diagram showing the pad plane, anode wires, shield and gated grid wires in the STAR MWPCs. The anode wires pitch is 4 mm, the shield and gated grid wires pitch is 1 mm. The distance from wires frame to pad is 2,4 and 10mm respectively.}
\label{fig:padplane}
\end{center}
\end{figure}

The iTPC upgrade project\cite{ref:note} will replace all 24 existing inner sectors in the STAR TPC with new, fully instrumented, sectors. The TPC and the iTPC upgrade use MWPCs with pad plane readout to record tracks of ionizing particles. There are three layers of wires above the pad plane: the anode wires (20 $\upmu$m diameter gold-plated tungsten wires), the shield wires and the gated grid (75 $\upmu$m diameter gold-plated beryllium copper wires), which are shown in Figure~\ref{fig:padplane}. The distance between the pad plane, the anode wire plane, the shield wire plane and the gated grid are 2 mm, 2 mm, and 6 mm respectively. The pitch for the anode wires is 4mm and 1 mm for the shield and gated grid wires.\\

For the MWPC construction, the wire tension must be strictly controlled as it is critical for the uniform gain performance of the TPC. The wires were wound first on wire-transfer frames and the wire tension was provided by the wire winding machine. The wire tension was then checked before a wire plane was transferred to the side wire mounts mounted on the individual sector strongbacks, using precision wire combs to secure the wire pitch and relative height of each wire plane\cite{ref:detector_3}. The wire tension was verified using the wire tension measurement system described in this paper. The wire tension will be checked both on the wire frame and also on the strongback.

\section{ The wire tension measurement method}
Various methods have been developed to measure the wire tension in the past years\cite{ref:detector_7,ref:detector_8,ref:detector_9}, which are generally based on electromechanical excitation, for example in a magnetic field\cite{ref:detector_7} or through capacitive coupling\cite{ref:detector_8}. In both cases, it requires electrical connection to wires, which is quite time-consuming, for a large amount of wires to be tested like our iTPC project. Each iTPC MWPC consists of more than 1500 wires. Here we use a laser-based optical system as described in below to measure the vibration induced by a short compressed gas jet, which does not need to touch the wire and can be done in 3\~{}4 seconds for one wire.\\

The tension T of a wire is related to its fundamental vibration frequency $f_{0}$ as in the following equation:
\begin{equation}
\begin{array}{l}
T=4 \mu f_0^2L^2  ,
\label{eq:1}
\end{array}
\end{equation}
where $\mu$ is the linear mass density, and L is the wire length between two fixed ends. If the fundamental frequency of a wire is known, then the wire tension can be calculated through Eq.\ref{eq:1}.\\

\begin{figure}
\begin{center}
\includegraphics[width=0.42\textwidth]{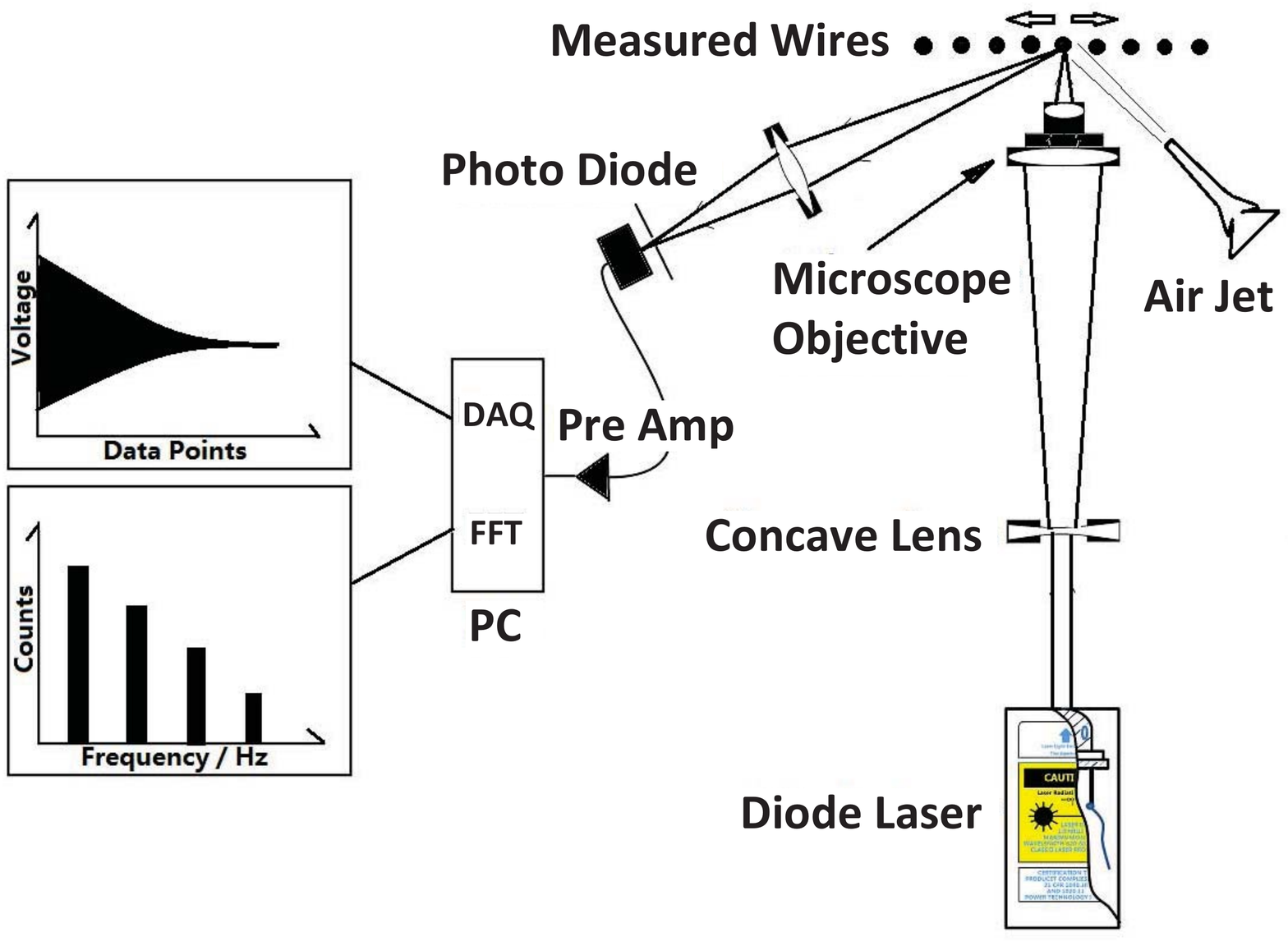}
\caption{A schematic diagram of the diode-laser based optical system which was used to make the wire tension measurements.}
\label{fig:laser_platform}
\end{center}
\end{figure}

Therefore, the key point for measurement is to find a way to transform the mechanical vibration of a wire to another signal that easily be detected. A laser based optical platform was designed to measure the vibration frequency of the wire as shown in Figure~\ref{fig:laser_platform}. The wires will vibrate under excitation of an air jet. When the air jet stops blowing, the wire will vibrate freely and the fundamental frequency of its vibration is an indication of the tension on the wire.\\

The wire is illuminated by a laser beam and the reflected light is collected by a photodiode, which can transform the light to analog signals. The frequency of the mechanical vibration of wire is equal to the reflected laser intensity fluctuation frequency. \\

Finally, the analog signals are digitized and recorded by a data acquisition system. The digitized signals are time domain data and they are transformed into frequency domain by means of the Fast Fourier Transform (FFT) algorithms. Consequently, the fundamental frequency is extracted from the frequency domain signals and the wire tension is calculated through Eq.\ref{eq:1} with parameters of linear density and wire length.\\


Sampling and digitization of the analog signals was done at the rate of 10 KS/s and the data length is 10 K sample points. The sampling rate is double the maximum frequency to be detected. According to the Nyquist theorem, the sampling rate is sufficient for accurate measurements.

\section{System setup}

\nopagebreak
\begin{figure*}
\begin{center}
\includegraphics[width=0.8\textwidth]{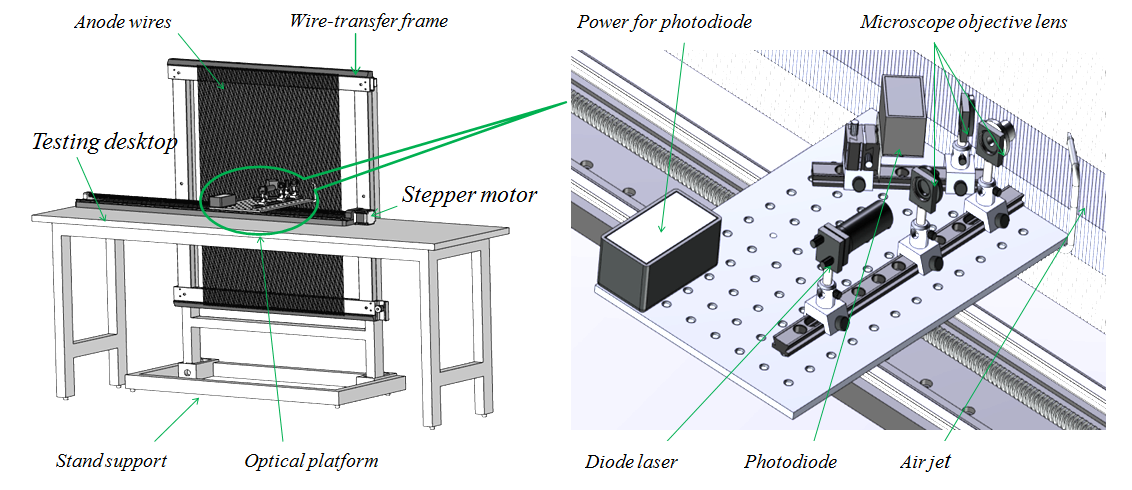}
\caption{The measurement bench (left) and the laser based optical platform (right).}
\label{fig:hardware_component}
\end{center}
\end{figure*}

Our wire tension measurement system can be used to measure multiple wires on a wire-transfer frame in a single run. The hardware and the software was designed to be fully automated.\\


\subsection{Hardware components}

Besides the laser based optical platform and the air jet, the hardware also includes a stepper motor. As shown in Figure~\ref{fig:hardware_component} (left), both the optical platform and the air jet were mounted on the stepper motor platform and can move with the motion of the platform. The wire transfer frames were placed on a support stand that is absolutely parallel to the moving direction of the stepper motor. Therefore, the laser beam can scan the wires on the frame one by one, and the tension of each wire was obtained.\\


The laser based optical platform and air jet is shown in Figure~\ref{fig:hardware_component}. When a measurement starts, the laser beam focuses on the wire to be detected. To get enough intensity of reflected image, the laser spot that is focussed on the wire must be approximately the same diameter as the wire. The air jet is driven by a pneumatic pump and it can blow directly upon the wire to be tested. In addition, the time duration and physical strength of the air jet can be adjusted to excite a suitable wire vibration.\\

To achieve automation for the laser beam focusing system, the motor's steps consists of coarse step and fine steps. The travel length of the coarse step is slightly smaller than the wire pitch. Data acquisition and digitalization are accomplished with the fine steps that are used to  determine the focal point for the laser beam.

\subsection{Software framework}

\begin{figure}
\begin{center}
\includegraphics[width=0.45\textwidth]{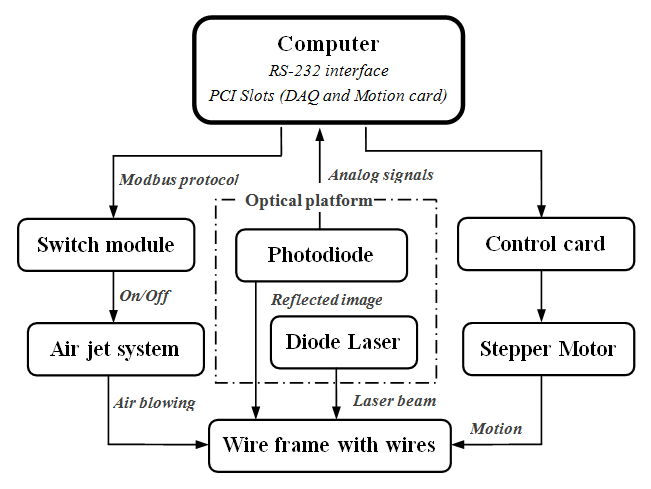}
\caption{Block diagram of the control scheme for the wire tension scanning system.}
\label{fig:software}
\end{center}
\end{figure}

The software was developed using the LabVIEW graphical programming environment on a Windows PC. The PC hosts a PCI bus, RS-232 serial interfaces, a motion control card (DMC2410) and a data acquisition PCI card (NI 6230: 16-Bit, 250 kS/s). It works as the  workbench to perform instruments control and monitoring, data acquisition and data processing as shown in Figure~\ref{fig:software}.\\

\begin{figure}
\begin{center}
\includegraphics[width=0.46\textwidth]{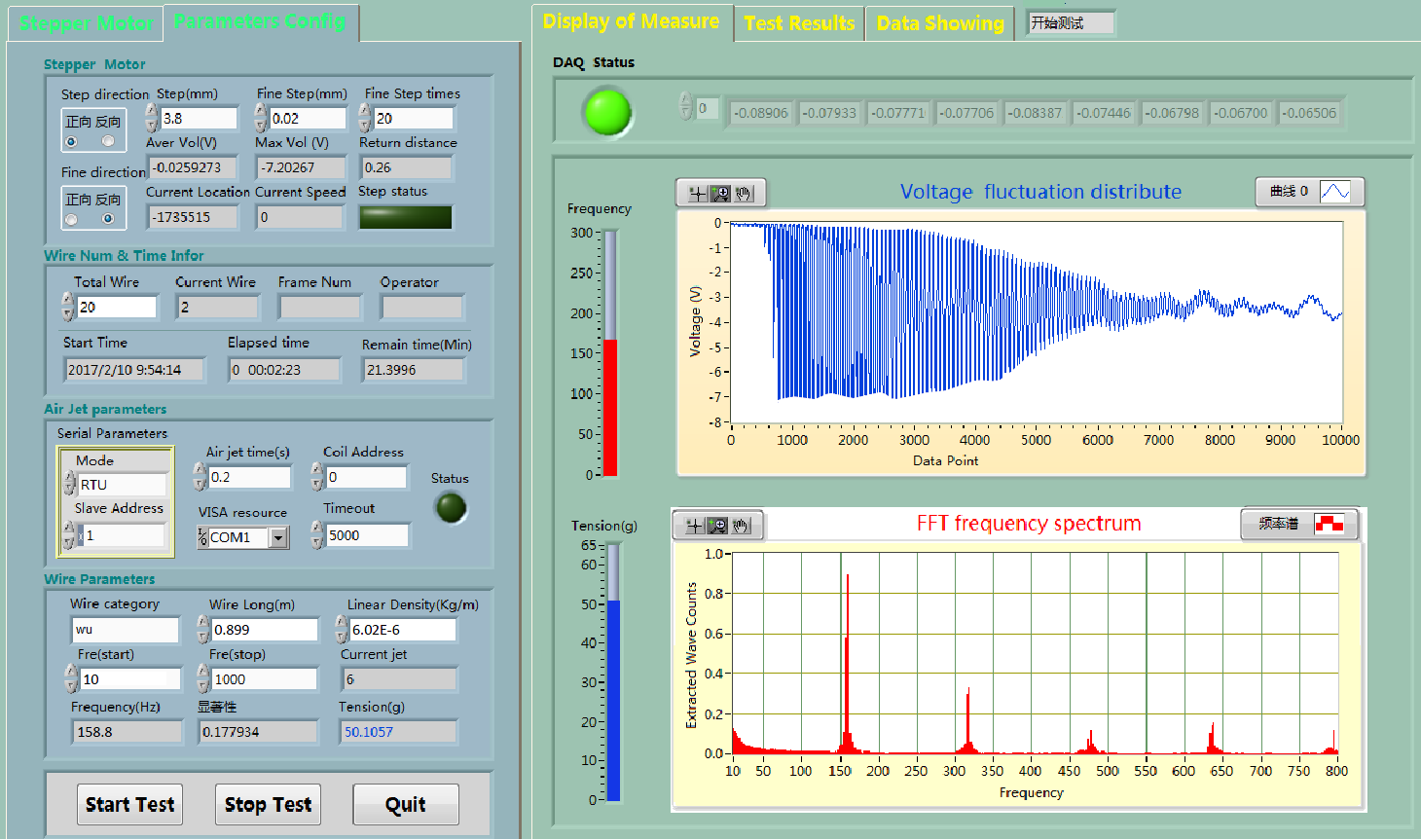}
\caption{Software graphical interface for wire tension measurement. The left part is used to input parameters and right part shows the measurement results.}
\label{fig:software_gui}
\end{center}
\end{figure}

The motion control card is responsible for the drive and control of the stepper motor. A serial interface is used to communicate with an intelligent switch module (MR-D0808-KN) using the Modbus protocol, which is used to activate a configurable set of power relays for turning the air jet on or off.\\

Sample and digitalization are fulfilled using a National Instruments NI 6230.  Data analysis is done by means of an FFT algorithm. A web page was built by HTML and PHP for the displaying real-time results on the web from the internet for remote users.\\

A graphical control interface for tension measurement was developed, as shown in Figure ~\ref{fig:software_gui}.  Columns are used to input the parameters, such as wire length, linear density, \textit{etc}, and it also shows the latest results. The graphical interface also features a set of status LEDs to show work status for each module of the system.

\section{Measurement results}

The wire tension was first measured on the wire-transfer frame. There are one hundred and sixty 20 $\upmu$m diameter gold-plated tungsten wires on the wire-transfer frame. The wire length is 0.899 $m$ and the linear mass density is 6.02 $\times$$10^{-6}$ $kg/m$. The tension was set to 50 grams during the wire winding operation.\\

\begin{figure}
\begin{center}
\includegraphics[width=0.46\textwidth]{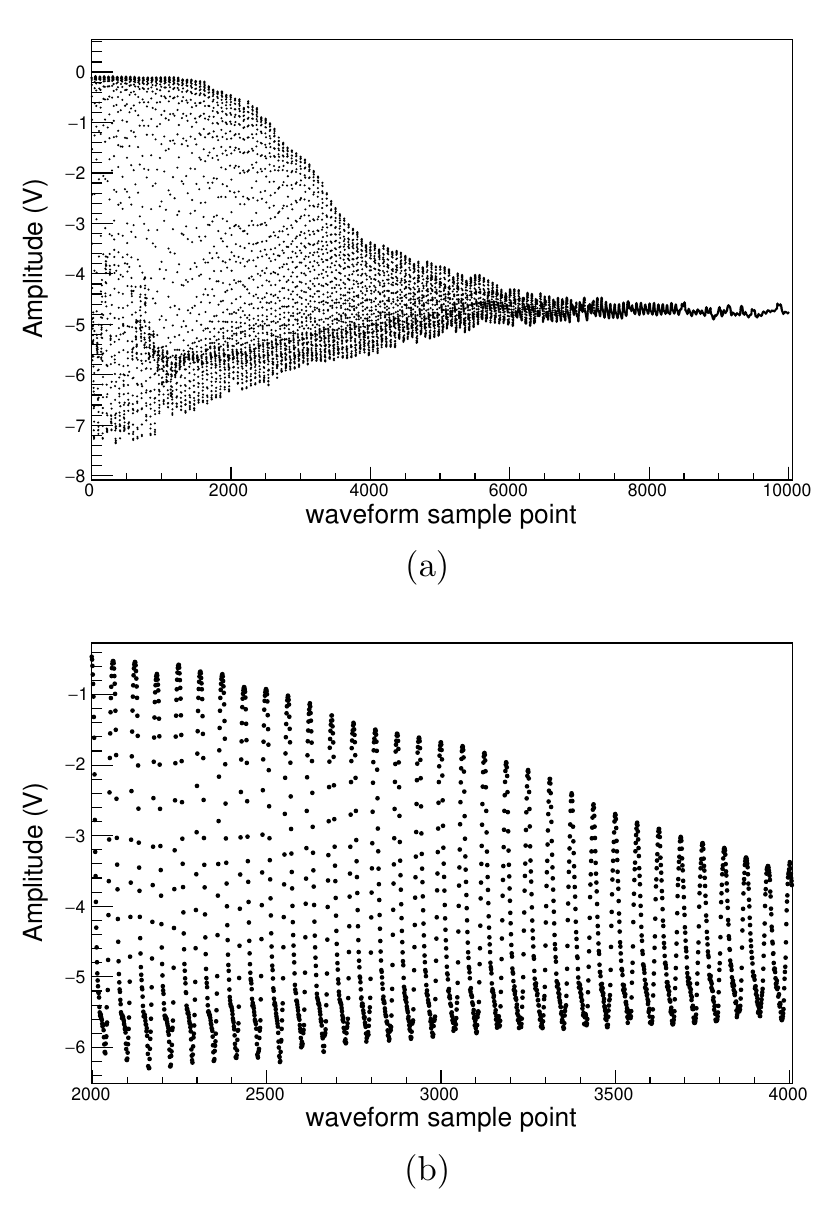}
\caption{The time domain signal acquired by the data acquisition system. The top plot (a) shows all of the sampling points acquired in one second and the bottom figure (b) is an expanded view showing the sample points from 2000 to 4000.}
\label{fig:fft_1}
\end{center}
\end{figure}

\begin{figure}
\begin{center}
\includegraphics[width=0.46\textwidth]{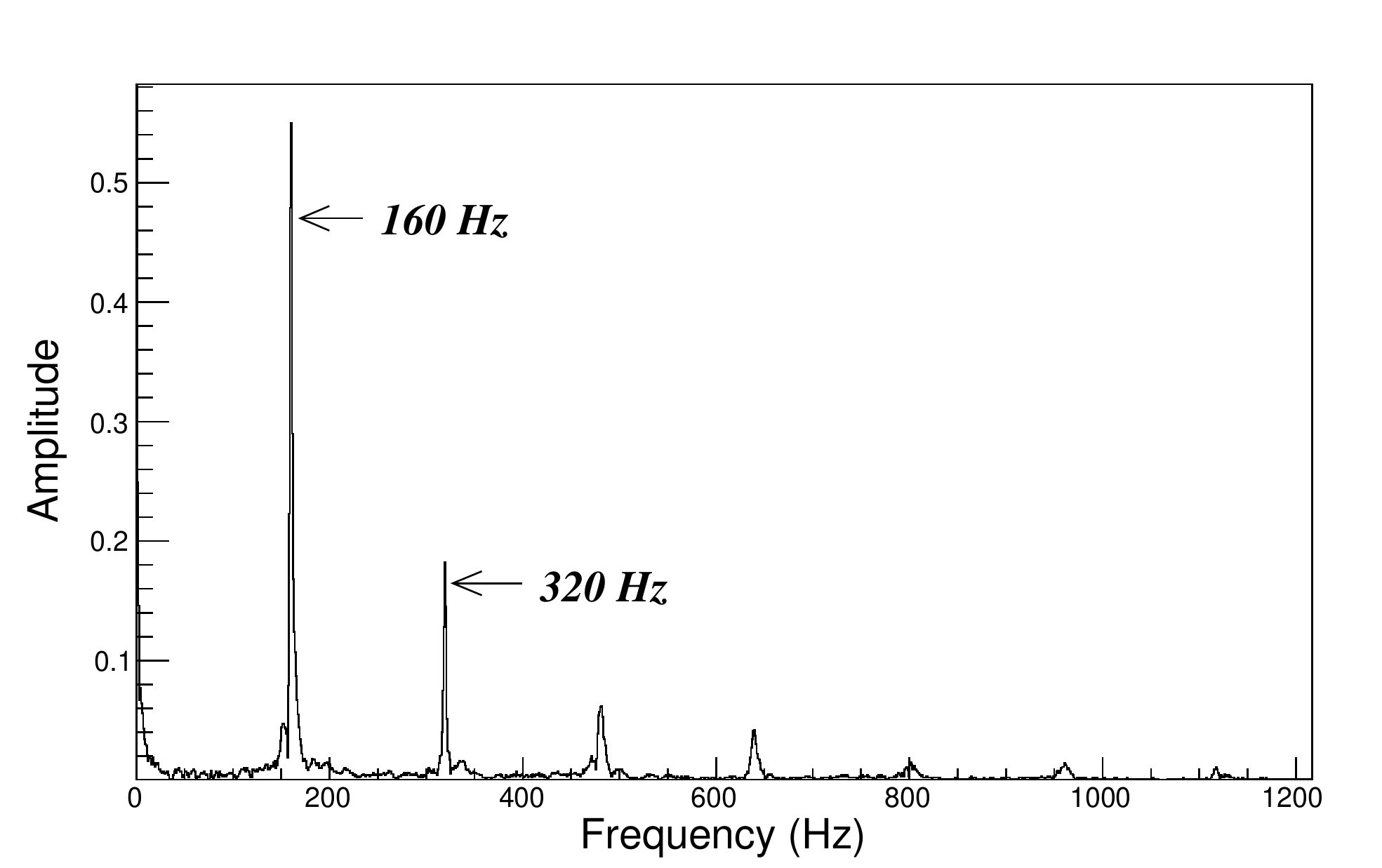}
\caption{The frequency spectrum transformed from the time domain by means of FFT. The first peak is the fundamental frequency and the $2^{nd}$ peak is twice the fundamental frequency, etc.}
\label{fig:fft_2}
\end{center}
\end{figure}

\begin{figure}
\begin{center}
\includegraphics[width=0.46\textwidth]{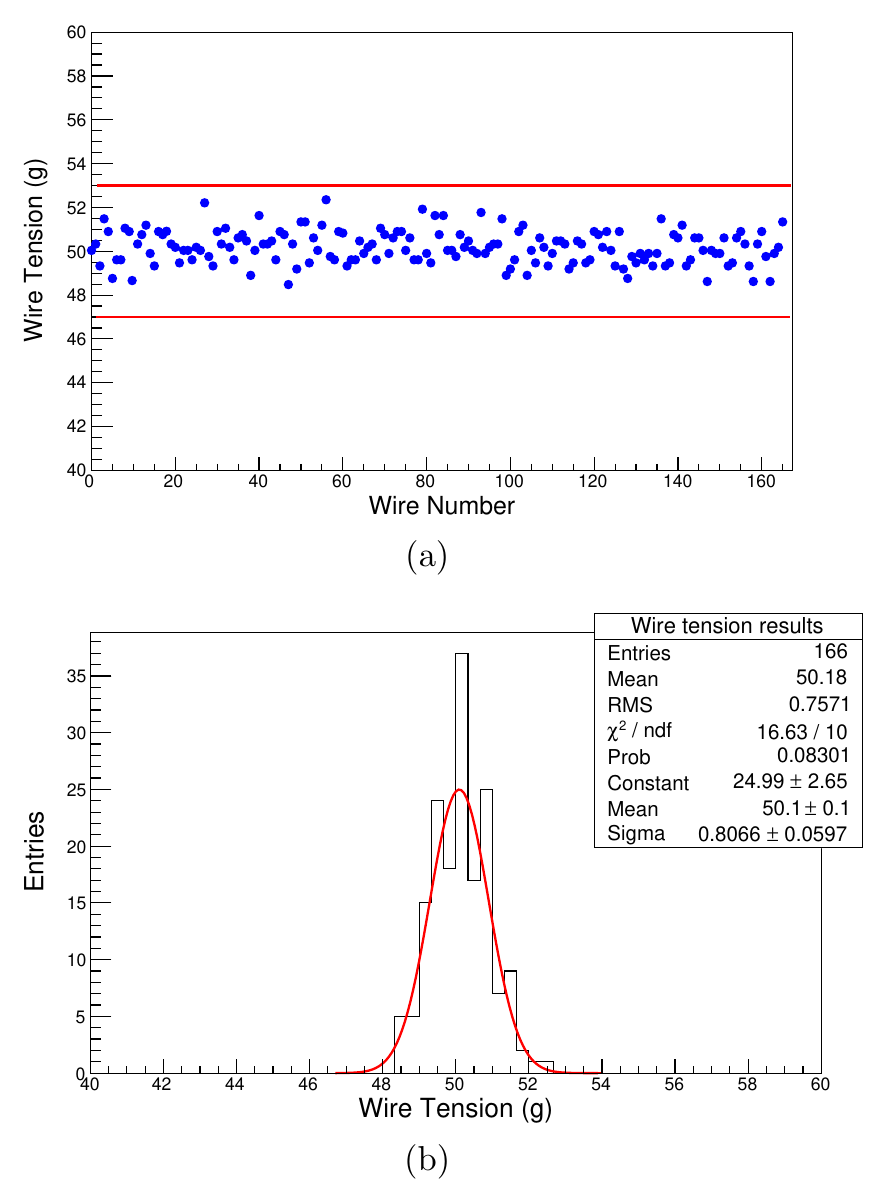}
\caption{The top plot (a) shows the wire tension measured on each wire on a wire transfer frame while the lower plot, the two lines indicate the acceptable range for iTPC upgrade (b) shows the distribution and variation of the results.}
\label{fig:tension}
\end{center}
\end{figure}

Figure~\ref{fig:fft_1} shows the time domain signal measured on one wire. Figure~\ref{fig:fft_2} shows the frequency domain spectrum transformed from the time domain using an FFT algorithm. The first peak in Figure~\ref{fig:fft_2} is the fundamental frequency and the second peak is twice the fundamental frequency, etc. The fundamental frequency, in this case, is 160 Hz. Figure~\ref{fig:tension} (a) shows the tension measurements for all of the wires on one transfer frame. The mean value for these measurements was 50.1 grams as shown in Figure~\ref{fig:tension} (b).  These results indicate that wire winding works well and the tension on all of the wires is uniform within an uncertainty of 0.8 gram(1.6\%). This is well within the acceptable range of 6\% uncertainty required by iTPC upgrade.\\

\section{Calibration and verification of the test system}

\begin{figure}
\begin{center}
\includegraphics[width=0.46\textwidth]{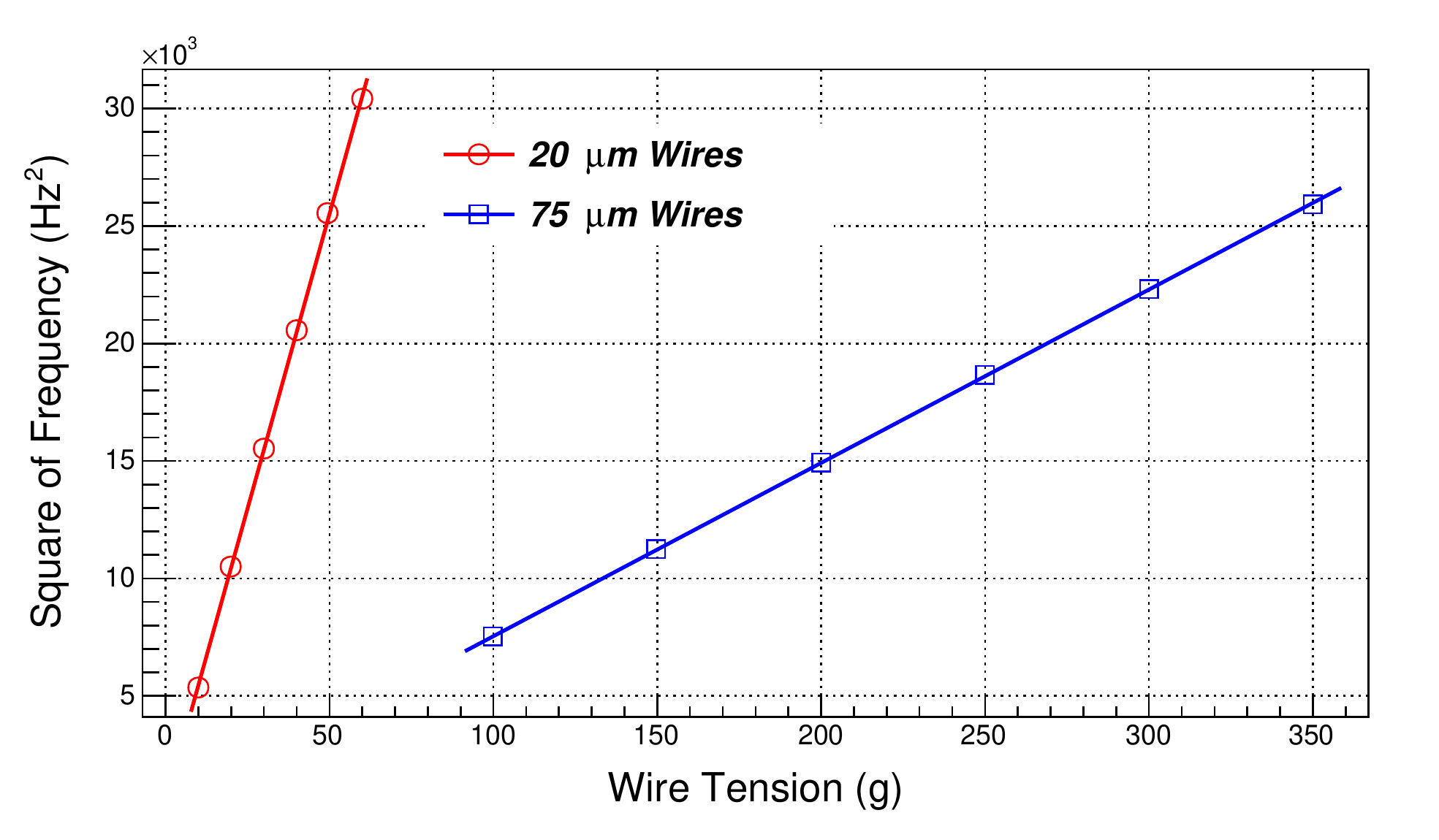}
\caption{The linearity between the square of the measured frequency and the wire tension for both 20 $\upmu$m and 75 $\upmu$m diameter wires. Note that the error bar is invisible as they are smaller than the marker size.}
\label{fig:verification_1}
\end{center}
\end{figure}

The calibration of the tension measurement system was done using one 20 $\upmu$m diameter gold-plated tungsten wire and one 75 $\upmu$m diameter gold-plated beryllium copper wire. Their lengths are the same, i.e., 0.899m and the mass density for beryllium copper wire is 4.06 $\times$$10^{-5}$ $kg/m$. Wires with known tension were prepared using standard weights. One end of the wire was soldered on the top-side of the wire frame and the other end was stretched with a standard weight. The fundamental frequency was extracted and the measured tension was compared with the standard weights. The calibration results show that both the 20 $\upmu$m and 75 $\upmu$m wires exhibit good linearity between the square of the measured fundamental frequency and the known wire tension as shown in Figure~\ref{fig:verification_1}. The error bar is estimated by repeating the measurements 100 times for each input tension and the RMS of the measurements is taken as the error bar, which are found to be very small ($<$0.8\%).
The plot indicates good linearity from 10 grams to 60 grams for 20$\upmu$m wires and 100-350 grams for 75 $\upmu$m wires, which covers the tension range for wires used in the STAR iTPC upgrade(50-120grams). \\

\begin{figure}
\begin{center}
\includegraphics[width=0.46\textwidth]{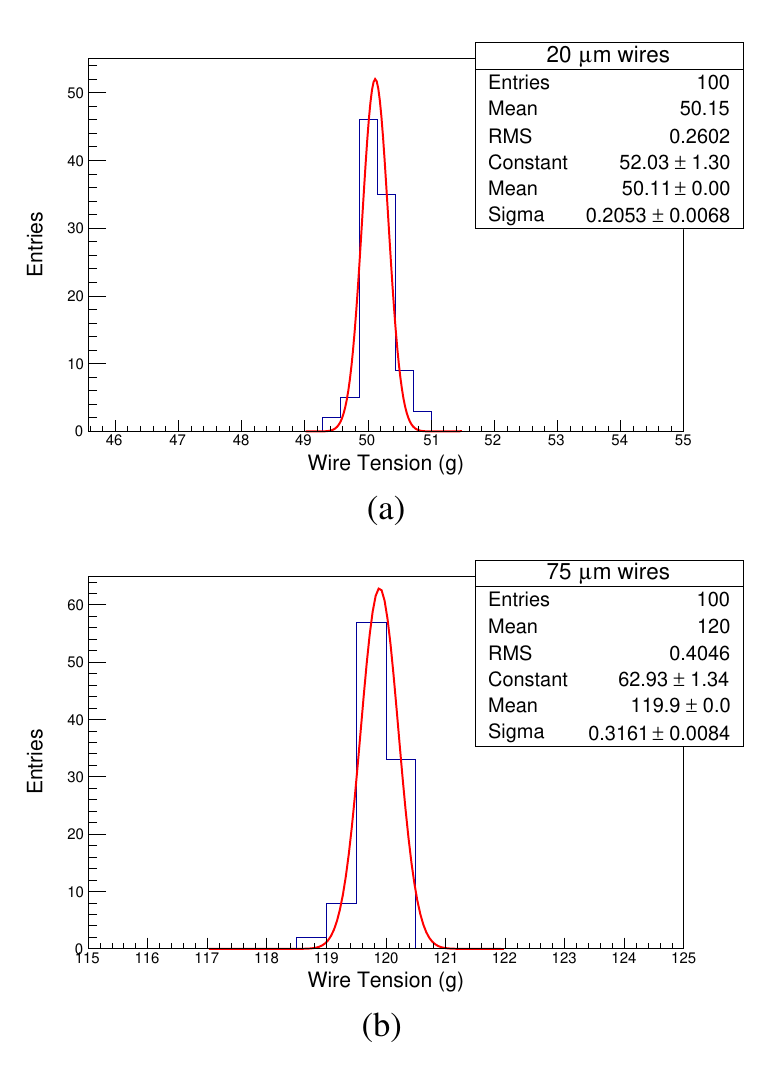}
\caption{The distribution of wire tensions measurements on calibrated wires. Two different wires and two different tension settings were investigated. Each wire was measured 100 times. The figure (a) is 20$\upmu$m diameter wire with 50 gram tension setting and (b) shows 75 $\upmu$m diameter wire applying 120 gram tension setting.}
\label{fig:verification_3}
\end{center}
\end{figure}

To verify the systematic uncertainty of the tension measurements, the tension was measured 100 times on the same wire for the two different diameter wires. Figure~\ref{fig:verification_3} (a) shows the measured results for 20$\upmu$m diameter wire with 50 gram tension and figure (b) shows the results for 75 $\upmu$m diameter wire with 120 gram tension. The RMS of these measurements are 0.26g and 0.40g, respectively.
The calculation uncertainties transferred from the errors of wire length (0.001m) and mass density ($10^{-8} kg/m$) are 0.14g and 0.27g for 20 $\upmu$m and 75 $\upmu$m wires respectively.
The total uncertainties (0.30g and 0.48g for wires with tension 50g and 120g respectively) are thus less than 1\% of the total tension on each wire.

\section{Conclusion}


A wire tension measurement system has been setup using a measurement method of detecting the fundamental frequency of wire vibration in MWPCs for the STAR iTPC upgrade project. The system can measure wire tension wire by wire, automatically, with good precision and stability. The system will be used to make wire tension measurements for all 24 MWPCs used in the STAR iTPC upgrade.

\section*{Acknowledgments}

We would like to thank the STAR collaboration for sharing some of the hardware and their strong encouragement. We would also like to thank the STAR iTPC upgrade collaboration members for their valuable suggestions and support. The work was supported in part by the National Natural Science Foundation of China (No. 11520101004), the Major State Basic Research Development Program in China (No. 2014CB845400), and the Natural Science Foundation of Shandong Province, China, under Grant No. ZR2013JQ001.


\section*{References}

\begin{thebibliography}{00}
\bibitem{ref:detector_1} The STAR Collaboration, The STAR Conceptual Design Report, June 15, 1992 LBLPUB-5347.
\bibitem{ref:detector_2} The STAR Collaboration, STAR Project CDR Update, Jan. 1993 LBL-PUB-5347 Rev.
\bibitem{ref:detector_3} H. Wieman, {\it et al.}, IEEE Trans. Nucl. Sci. 44 (1997) 671.
\bibitem{ref:detector_4} K. Ackermann, Nucl. Phys. A 661(1999) 686c.
\bibitem{ref:detector_5} M. Anderson {\it et al.}, Nucl. Instr. Meth. A 499 (2003) 659-678.
\bibitem{ref:note} Technical Design Report for the iTPC Upgrade, Star Note 0644, https://drupal.star.bnl.gov/STAR/starnotes/public/sn0644
\bibitem{ref:detector_7} M. Calvetti, et al.[UA1 Collaboration], Nucl. Instrum. Meth. 174 (1980) 285.
\bibitem{ref:detector_8} P. Ciambrone et al., Nucl. Instrum. Meth. A 545 (2005) 156.
\bibitem{ref:detector_9} S. Bhadra, S. Errede, L. Fishback, H. Keutelian and P. Schlabach, Nucl. Instrum. Meth. A 269 (1988) 33.
\end{thebibliography}
\end{document}